# Forecasting Behavioral Events from Dopamine-Circuit Activity: Temporal Hyperparameter Search

Sergey Grigorovich[1,2], Maryia Bairachnaya[1,2], Bruno Giros[1,2]

[1]Douglas Reseach Centre, Douglas Mental Health Institute, Montreal, QC, Canada
[2]McGill University, Montreal, Quebec, Canada
sergey.grigorovich@mail.mcgill.ca

**Abstract.** In the present study, we examine how choices of temporal hyperparameters affect forecasting performance in models that predict physiological processes. Our dataset consists of fiber-photometry recordings of GCaMP6 calcium signal from the population of dopaminoceptive neurons in the ventrolateral striatum, together with licking behavior in a head-fixed protocol. We use generalized linear models (GLMs) to predict individual licking events and burst onsets from the recorded neural signal while scanning a grid of history-window lengths and forecast horizons to evaluate cross-subject generalization. We compare the DA channel against an isosbestic baseline, to control for residual non-DA predictors. The DA signal shows the largest advantage for short-timescale single event forecasting, whereas burst-start prediction peaks at intermediate horizons and at both short and longer history windows. These results demonstrate how the selection of temporal hyperparameters depends on the behavioral target and underlying mechanisms.

**Keywords:** Behavioral event forecasting, Dopamine-behavior coupling, Parameter search

## 1 Introduction

In neuroscience, the analysis of the neural activity's link to behavior often boils down to a time-series prediction problem: the forecasting of specific behavioral events based on a recorded neural signal. It immediately raises more questions, particularly, how far in the future could we extend our predictions, and over which timescale does the neural signal carry useful information? The answer will vary depending on the measurement technique, the neural circuitry involved, the nature of the task in the experimental paradigm, etc.

Fiber photometry is an imaging technique that captures population-level calcium (Ca2+) activity from specific cell types within a brain region of interest. In combination with genetic tools to target specific populations, it has become a widely used approach for monitoring neuronal activity in vivo. However, fiber photometry data contain substantial non-neural variance from motion artifacts, optical effects, and hemodynamics [8], which affects interpretation and may inflate predictive performance.

In dopaminoceptive (DA) neural circuits, recorded signals combine both rapid and slow components that incorporate momentary action traces and reward expectation [4]. In addition to features of the underlying mechanisms, each attempt to link neuronal activity to observable behavior faces multiple practical questions. In the present study, we address two of them: (i) what temporal context length produces the best generalization to new subjects, and (ii) which behavioral target definition is most predictable from neural measurements - individual events or bursts/state transitions. These questions remain important because they can change an approach to “what is predictable” without changing the underlying data, simply by altering the temporal framing of the prediction task.

## 2 Methods

### 2.1 Dataset

**Animal subjects.** For the initial data acquisition, we used 9 genetically modified mice that produce Cre recombinase (Cre) in neurons expressing Dopamine type-1 receptor (D1R). The viral solution (titer ~1-2 x 10^12 vg/ml) was injected with a Glass Capillary Nanoinjector (Neurostar, Germany) with an attached glass capillary. A volume of 500 nL of AAV1-hSyn-DIO-GCaMP6f (#100833, Addgene) was unilaterally injected into the ventrolateral striatum (VLS, coordinates from bregma: AP +0.2 mm, ML +2.5 mm, DV −4.4 mm). Mice were allowed to recover for at least 21 days post-injection to ensure optimal AAV expression. VLS was chosen as a primary area of interest due to its high involvement in orofacial behaviors, specifically licking. The coordinates were chosen based on Bakhurin et al. [1]

**Dataset and experimental setup.** The dataset consist of the recordings from calcium imaging fiber photometry and a licking sensor, collected from 9 head-fixed, water-restricted mice across 30 behavioral sessions per mouse. The protocol generally follows the schedule introduced by Toda et al. [7]. Each session lasted 25-30 minutes. Rewards (2 µl of 10% sucrose solution) were delivered once per trial (every 10 s), without an explicit sensory cue. Alongside the calcium-dependent fluorescence channel (D1R-expressing neurons population signal), an isosbestic control channel was recorded for motion/artifact tracking. Both photometry channels and licking sensor were sampled at 20 Hz.

### 2.2 Data preparation

**Photometry preprocessing.** Raw fluorescence traces from the calcium-dependent 470 nm channel and the 415 nm isosbestic control channel were preprocessed to account for a photobleaching, suppress high-frequency noise, and reduce motion/optical artifacts. To remove a baseline drift associated with photobleaching, we fitted each channel with a bi-exponential decay model and subtracted the fitted trend from the corresponding raw trace, obtaining detrended 470 nm and 415 nm signals.

Both detrended signals were passed through a low-pass Butterworth filter (cutoff at 3 Hz, normalized to the 10 Hz Nyquist frequency) to reduce high-frequency measurement noise.

To account for motion/optical artifacts, we fitted the filtered 415 nm trace to the filtered 470 nm signal using a robust regression model with Huber loss. The choice of robust regression over ordinary least squares is based on its ability to reduce sensitivity to outliers and mitigate cases where control fitting absorbs neural transients [3]. Finally, we normalized both the detrended/filtered 415 nm signal and the corrected 470 nm signal using a median and median absolute deviation (MAD) z-score ("modified z-score") to reduce sensitivity to extreme values.

**Behavioral event extraction.** Licking events were represented as a point process from detected lick timestamps. We defined lick-related burst structure using an inter-lick interval (ILI) threshold of 250 ms [2]. A burst start event was defined as the first lick following an ILI > 250 ms, with at least 3 licks before ILI > 250 ms.

**Prediction targets.** We evaluated two target metrics at each time bin (t) in the photometry history to predict outcomes over a future forecast horizon (H):

1. Lick occurrence (binary): defined as 1 if one or more events are present in the interval (t; t+H], otherwise 0.

2. Burst transition occurrence (binary): defined as 1 if one or more licking burst state changes (see above) are present in the interval (t, t+H], otherwise 0.

Forecast horizons were drawn from a predefined grid H in {0.1, 0.25, 0.5, 1, 2} seconds (2–40 bins at 20 Hz).

**Photometry history representation (log-time raised-cosine basis).** The comparison of different historical window lengths (L) may introduce significant imbalance due to the variation of model features number. To keep the number of predictors constant, we represented photometry history using a fixed set of 5 temporal basis functions. For each history length (L), we constructed a log-time raised-cosine basis, using the time warp:

$$u(\tau)=\log(\tau+\epsilon)$$

Each basis function is defined as a compactly supported raised cosine in warped time, producing higher temporal resolution near $\tau = 0$ and progressively lower resolution at longer lags [6]. In this way, for each time bin, the historical features were computed as a causal convolution of the pre-processed photometry signal. This approach produces a design matrix with a constant number of predictors per session for any history length. We evaluated history lengths L in {0.25, 0.5, 1, 3, 5, 8} seconds.

## 2.3 Predictive models

To predict future outcomes from derived features, we used generalized linear models (GLMs). For both binary targets, we fit binomial logistic regression models, resulting in the estimation of the probability of event occurrence in the forecast horizon.

To control for non-stationary and known structure in the behavioral protocol, we included two additional covariates:

1. Session number, to account for gradual changes across training (task acquisition / learning progress).
2. Time within trial, to account for the cyclic structure of the task (10s trial).

These covariates prevent schedule-driven and training-driven variation from being attributed to neural inputs.

**Control analysis with isosbestic-only inputs.** As a control for residual motion/optical artifact-related effect, we repeated the same modeling pipeline using Iso z-scores as the input instead of the DA-associated calcium signal. This provides a benchmark for how much predictive performance can be achieved from motion artifacts only. Due to the nature of recorded motor activity (licking), we expect that a significant part of the observed behavior can be predicted from this channel alone.

**Evaluation metrics.** Due to the high volume of events in forecast windows, especially in the wider ones, we quantified model performance using Precision@1% (P@1%), a high-confidence metric that emphasizes the low-recall region of the precision–recall tradeoff [5]. Photometry signals can retain movement/hemodynamic components even after correction; thus we compared a model based on the preprocessed DA channel against an Iso-only control to estimate how much predictive performance could be achieved from artifact-related structure alone and evaluate gains of the DA model.

Thus, we included the following evaluation metrics in the report:

1. DA signal lift over random:

$$P@1\%_{DA}(L,H) - \pi(L,H)$$

2. Iso lift over random:

$$P@1\%_{Iso}(L,H) - \pi(L,H)$$

3. DA signal gain over Iso:

$$P@1\%_{DA}(L,H) - P@1\%_{Iso}(L,H)$$

where $\pi$ represents “random precision” (prevalence)

All analyses were performed in Python using scikit-learn, statsmodels, and SciPy libraries for data preprocessing, splits, model fit and evaluation. Random seeds were fixed for reproducibility. Hyperparameter grids were predefined as described above.

## 3 Results

We evaluated cross-subject generalization using leave-one-mouse-out (LOO) testing. For each LOO mouse and each combination of history window length (L) and forecast horizon (H), we computed predicted probabilities for both individual licking events and “start of the burst” events. These probabilities were compared to “random precision” $\pi$ at the same (L, H) to produce evaluation metrics (see Methods). “DA signal lift over random" and "Iso lift over random" provide important context and general insights about model performance. However, we consider "DA signal gain over Iso" the main metric of model's ability to capture prediction power of DA signal over other predictors.

### 3.1 Predicting individual licking events

For the single lick prediction, the DA model demonstrated that its advantage over the Iso baseline depends on the selected history/forecast window sizes. The largest gain occurs at short timescales, peaking at L=0.5 s and H=0.25 s, with a mean gain of ΔP@1% = 0.148 and SD = 0.05 across LOO mice. For larger temporal windows (L > 0.5s and H > 0.5s), the DA–Iso gain diminishes, remaining in the range ΔP@1% ≈ 0.02–0.08 (see Fig. 1). To estimate gains over label prevalence, see Table 1.

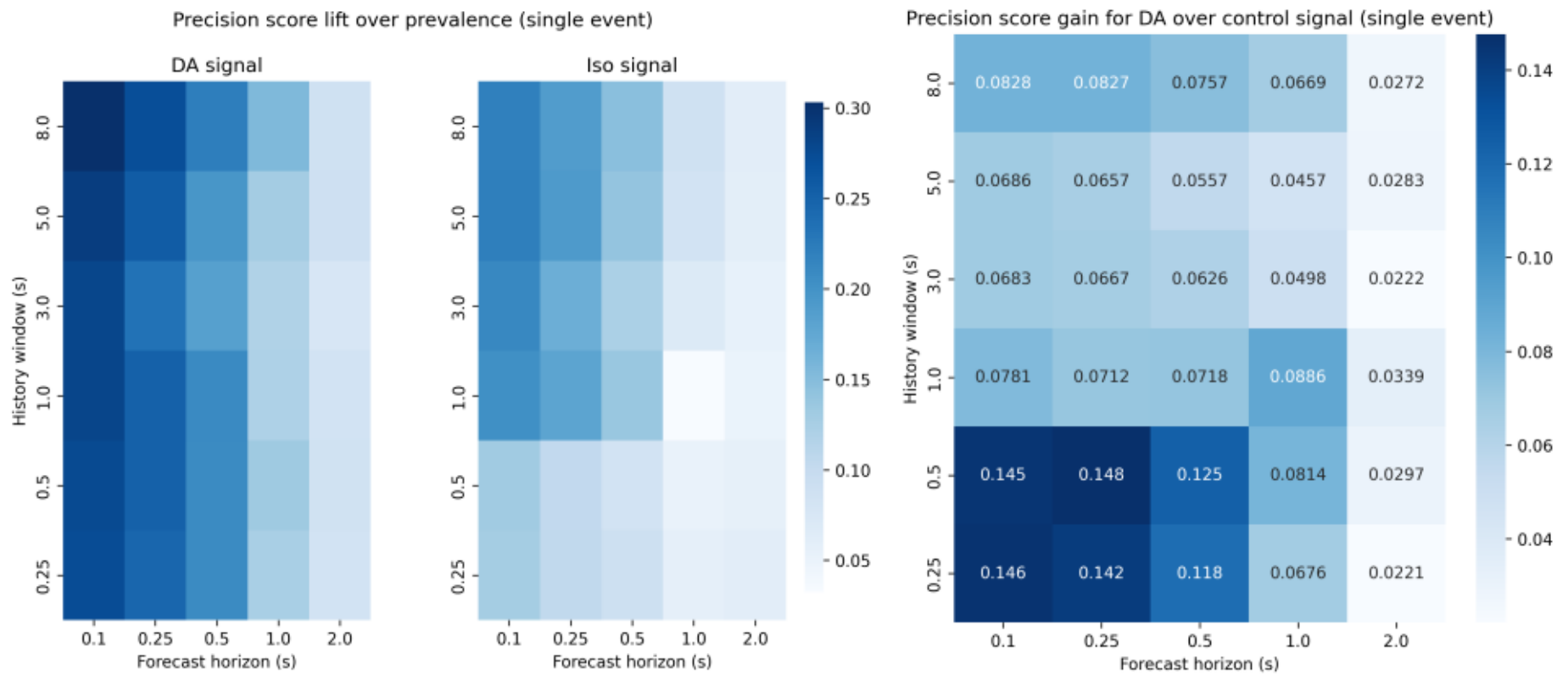

**Fig. 1**. High-confidence forecasting for single-lick events. Heatmaps show performance across history window L (y-axis) and forecast horizon H (x-axis) using P@1%, evaluated on LOO mice. Left: DA model lift over prevalence. Middle: lift for the Iso-only control model. Right: DA gain over control, higher values indicate more accurate high-confidence prediction. For random prevalence over H see Table 1.

Table 1. Label prevalence by forecast horizon. Baseline (random) precision π for each forecast horizon H for the single-lick and burst-start labels. Values represent mean precision π across LOO mice, values in parentheses indicate the standard deviation.

| H | 0.1s | 0.25s | 0.5s | 1s | 2s |
|---|---|---|---|---|---|
| Single event | 0.23 (0.02) | 0.34 (0.03) | 0.46 (0.04) | 0.62 (0.04) | 0.81 (0.03) |
| Burst start | 0.02 (<0.01) | 0.05 (<0.01) | 0.09 (0.01) | 0.18 (0.01) | 0.33 (0.02) |

These results indicate that the recorded photometry data from the DA neurons contains additional short-timescale predictive information that improves high-confidence forecasting of upcoming immediate lick events beyond what is captured by the Iso control. It can be interpreted as evidence of the correlation of the fast dynamics in the DA channel with near-future event likelihood. It is essential to note that calcium imaging photometry signals can still reflect mixtures of neural and residual movement artifact components, which weren't captured by Iso channel fitting.

### 3.2 Predicting burst-start events

Compared to individual lick event prediction, burst-start prediction exhibited a different structure in the (L, H) performance landscape. Specifically, the DA-over-Iso gain demonstrates two areas of elevated performance, both centered on H=1s: at L=1s and at L=8s (see Fig. 2).

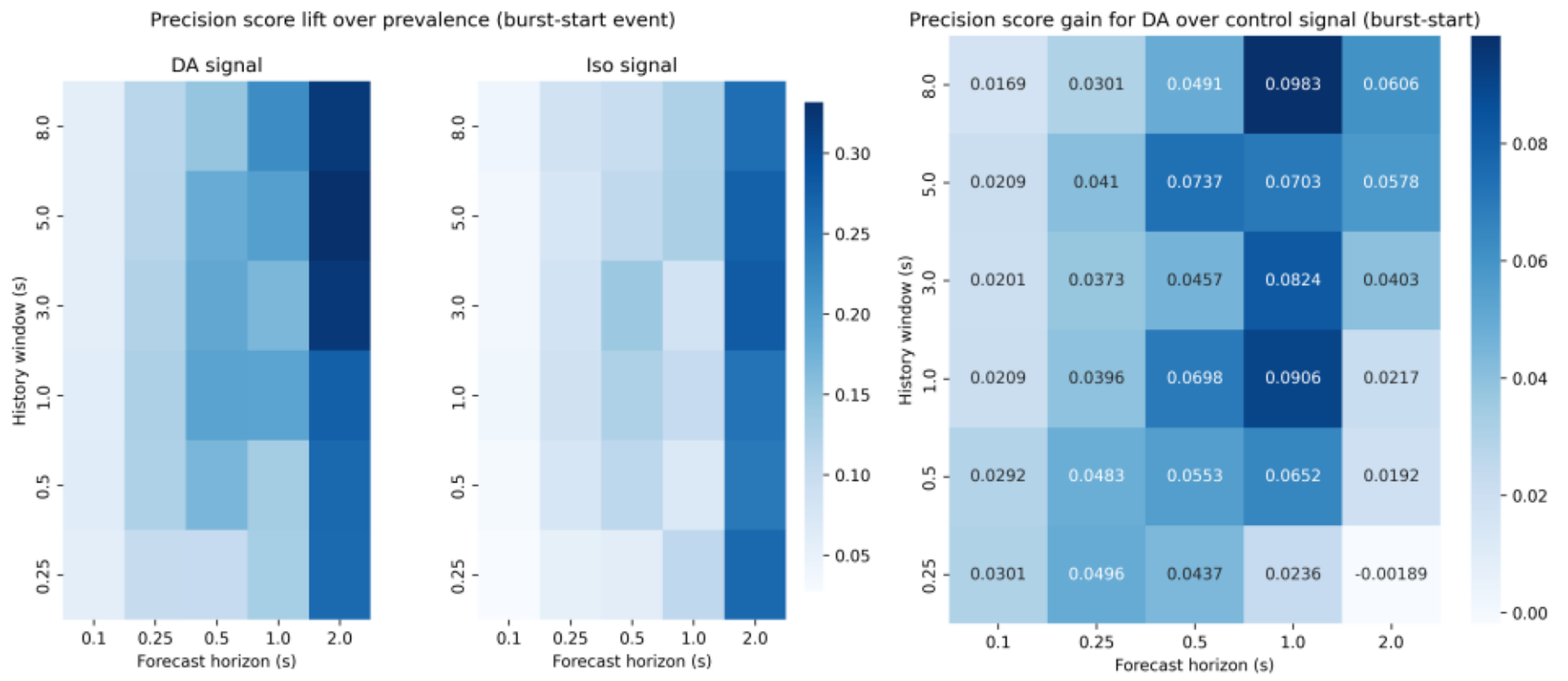

**Fig. 2**. High-confidence forecasting for burst-start events. Heatmaps show performance across history window L (y-axis) and forecast horizon H (x-axis) using P@1%, evaluated on LOO mice. Left: DA model lift over prevalence. Middle: lift for the Iso-only control model. Right: DA gain over control. For random prevalence over H see Table 1.

A plausible explanation for the L=8s region follows from the fixed structure of the experiment. Rewards were delivered every 10s without an explicit cue, and head-fixed mice typically show a stereotyped cycle consisting of the post-reward consumption, followed by post-consumption pause, and then rising anticipatory licking as the next reward approaches (for well-trained animals). Under this schedule, an 8–9 s history window can capture a substantial fraction of any slowly decaying component associated with the previous consumption episode. The long-history peak may reflect schedule-aligned dynamics that extend over several seconds, rather than the benefits of wider history windows.

For the medium windows peak (H=1s, L=1s), we suggest an explanation that burst starts are best predicted at a timescale that matches the underlying natural structure of licking patterns. In mice, licking often occurs in short bursts (on the order of 5–12 Hz, with 3-8 licks per burst), so, with a 1s horizon, model effectively forecasts whether a transition into a lick bout will occur soon. On the other hand, a 1s history window provides enough context to capture upcoming state changes, while remaining short enough to remain relevant for event prediction.

### 3.3 Summary

For the prediction of both individual licking events and burst starts, the DA channel shows a timescale-dependent advantage over the Iso control. For individual lick-event forecasting, the DA gain was the largest for very short history/horizon windows. This indicates that DA provides a relevant fast predictive component for near-future event prediction.

For burst-start forecasting, the events can be predicted with higher confidence within a wider forecast horizon, which reflects the nature of the underlying neuronal mechanism. The shift toward an intermediate (1s) forecast window can be partially explained by the accumulation of released DA in downstream licking pattern generators. These results support the contributions of both intermediate history and long history activity. However, for our behavioral task, the long-history region can be explained by the periodic reward schedule and the behavioral cycle of post-consumption. With no such explanation for the intermediate-history region, we conclude that it is likely associated with actual consumption and/or anticipatory licking.

## 4 Conclusion

In the present study, we have demonstrated that DA neurons activity, recorded by calcium imaging fiber photometry, provides predictive information beyond the Iso control. This advantage is not consistent across the history/forecast window (L, H) space and varies for different metrics, making the selection of model hyperparameters even more challenging.

For individual lick events, the DA channel's predictive power is largest for very short history/horizon windows, consistent with the rapid nature of single movement actions. For burst-start prediction, the DA channel advantage is observed around an intermediate horizon. It exhibits a more complex structure with peaks for intermediate- and long-history windows. We hypothesize that this reflects the combined influence of state transitions and longer, schedule-aligned dynamics in fixed-interval reward protocols.

These findings suggest the importance of alignment between evaluation metrics and the timescale of the studied phenomena. For protocols with slower dynamics (such as the present project, reward anticipation design), metrics and label definitions should emphasize state transitions and rely on wider historical windows and longer horizons. In contrast, experimental protocols focused on fast stimulus–response events may benefit from single event-associated metrics with immediate history and short-horizon forecasting. Thus, selecting optimal metrics and horizons/history lengths is not a technical detail. These choices reflect the hypothesis about the dominant temporal scale of the underlying mechanisms.

**Disclosure of Interests.** The authors have no competing interests to declare that are relevant to the content of this article.